\documentclass[12pt]{article}
\usepackage{epsfig}

\date{\today}

\usepackage{amsmath}
\usepackage{amssymb}
\usepackage[hang,nooneline,scriptsize]{subfigure}
\renewcommand{\thefootnote}{\arabic{footnote}}

\makeatletter
\let\@fnsymbol\@arabic
\makeatother

\begin{document}
\title{Black Holes with Scalar Hairs in Einstein-Gauss-Bonnet Gravity}

\author{
{\large Y. Brihaye}$^{\dagger}$\footnote{email: yves.brihaye@umons.ac.be} , 
{\large L. Ducobu }$^{\dagger}$\footnote{email: ludovic.ducobu@student.umons.ac.be}
\\ \\
$^{\dagger}${\small Physique-Math\'ematique, Universit\'e de
Mons, 7000 Mons, Belgium}\\ 
}


\date{\today}
\setlength{\footnotesep}{0.5\footnotesep}
\newcommand{\dd}{\mbox{d}}
\newcommand{\tr}{\mbox{tr}}
\newcommand{\la}{\lambda}
\newcommand{\ka}{\kappa}
\newcommand{\f}{\phi}
\newcommand{\vf}{\varphi}
\newcommand{\F}{\Phi}
\newcommand{\al}{\alpha}
\newcommand{\ga}{\gamma}
\newcommand{\de}{\delta}
\newcommand{\si}{\sigma}
\newcommand{\bomega}{\mbox{\boldmath $\omega$}}
\newcommand{\bsi}{\mbox{\boldmath $\sigma$}}
\newcommand{\bchi}{\mbox{\boldmath $\chi$}}
\newcommand{\bal}{\mbox{\boldmath $\alpha$}}
\newcommand{\bpsi}{\mbox{\boldmath $\psi$}}
\newcommand{\brho}{\mbox{\boldmath $\varrho$}}
\newcommand{\beps}{\mbox{\boldmath $\varepsilon$}}
\newcommand{\bxi}{\mbox{\boldmath $\xi$}}
\newcommand{\bbeta}{\mbox{\boldmath $\beta$}}
\newcommand{\ee}{\end{equation}}
\newcommand{\eea}{\end{eqnarray}}
\newcommand{\be}{\begin{equation}}
\newcommand{\bea}{\begin{eqnarray}}

\newcommand{\ii}{\mbox{i}}
\newcommand{\e}{\mbox{e}}
\newcommand{\pa}{\partial}
\newcommand{\Om}{\Omega}
\newcommand{\vep}{\varepsilon}
\newcommand{\bfph}{{\bf \phi}}
\newcommand{\lm}{\lambda}
\def\theequation{\arabic{equation}}
\renewcommand{\thefootnote}{\fnsymbol{footnote}}
\newcommand{\re}[1]{(\ref{#1})}
\newcommand{\R}{{\rm I \hspace{-0.52ex} R}}
\newcommand{\N}{{\sf N\hspace*{-1.0ex}\rule{0.15ex}%
{1.3ex}\hspace*{1.0ex}}}
\newcommand{\Q}{{\sf Q\hspace*{-1.1ex}\rule{0.15ex}%
{1.5ex}\hspace*{1.1ex}}}
\newcommand{\C}{{\sf C\hspace*{-0.9ex}\rule{0.15ex}%
{1.3ex}\hspace*{0.9ex}}}
\newcommand{\eins}{1\hspace{-0.56ex}{\rm I}}
\renewcommand{\thefootnote}{\arabic{footnote}}
 \maketitle
\begin{abstract} 
The Einstein-Gauss-Bonnet gravity in five dimensions is extended by scalar fields and the corresponding
equations are  reduced to a system of non-linear differential equations. 
A large family of regular solutions of these equations is shown to exist. Generically, these solutions are
spinning black holes with scalar hairs. 
They can be characterized (but not uniquely) by an horizon and an angular velocity on this horizon.
Taking particular limits the black holes approach boson star or become extremal, in any case the limiting configurations
remain hairy.   
\end{abstract}
\medskip 
\medskip
 \ \ \ PACS Numbers: 04.70.-s,  04.50.Gh, 11.25.Tq

\section{Introduction}
Much attention was devoted during the recent years to gravity theories supplemented by scalar fields.
There are several reasons for that~: scalar fields appear as fundamental constituents in 
the standard model of particle physics as well as in numerous models attempting to go beyond this well tested theory.
Bypassing the long standing 'No Hair Theorem', black holes solutions with scalar hairs have been constructed 
in  \cite{Herdeiro:2014goa}, \cite{Herdeiro:2015gia}. These new solutions
of the Einstein-Klein-Gordon equations were constructed for a single massive complex
scalar field and exist provide the black hole rotates sufficiently fast;  as so they provide
hairy  generalizations of Kerr black holes. 
The influence of a self-gravitating potential on these solutions was studied in \cite{Herdeiro:2015tia}.

Among the numerous theoretical ideas explored to encompass the standard model and, hopefully, to  improve our
 understanding of the Universe, models formulated in  more than four dimensional space-times 
 emerge as promising candidates. Progressively they motivate the study of general relativity in $d$ dimensions.
 The understanding of the gravitational interaction in higher dimensions and the classification of the underlying
 classical solutions, like black holes and other compact objects, constitue a research topic by itself. 
 Even in the vacuum,
 the richness of the solutions available in $d$ dimensions \cite{myers_perry} is remarkable.
 In particular the black holes are characterized by $[(d-1)/2]$ independent angular momenta.
  For odd number of dimensions, spinning solutions
 with all equal angular momentum present an enhanced symmetry, the underlying equations can be set
 in  co-dimension one problem allowing  for  simplifications in the more elaborated dynamical equations. 
  This feature can be used,  in some circumstances, to study some dynamical processes,
  (e.g.   collapsing shells in rotations \cite{Delsate:2014iia}).
  Some calculations can be performed more easily in a space-time with five -rather than four-
 dimensions. 
  
 The construction of hairy black holes in higher dimensions therefore appeared as a natural continuation
 of the work  \cite{Herdeiro:2014goa} and it was the object of    \cite{Brihaye:2014nba}. In that paper
 it was shown that Myers-Perry black holes with scalar hairs exist on a specific domain of the two-dimensional
 parameter space of the problem~: the black hole horizon and the angular velocity of the solution on the 
 horizon. Interestingly, and contrasting with respect to 
 their 4-dimensional analogs, these hairy black holes do not approach the
 pure vacuum Myers-Perry solutions in any limit. The two  families are disjoined, separated  by a mass gap. 
 This property provides an
 example of difference of the gravitational interaction which can occur for different dimensions. Such 
 differences can be accounted on the  non-linear character of the Einstein equations.
 
 The black holes obtained in \cite{Brihaye:2014nba} and in this paper are asymptotically flat
 and characterized by the fact that
 they possess a single Killing field. This notion was introduced in \cite{Dias:2011at}
 where families of asymptotically AdS, spinning, hairy black holes and bosons stars were also constructed in
 five dimensions. In this case, pure AdS space-time is approached uniformly while the scalar field tends to zero.
 The generalization of these solutions to higher dimensions is emphasized in \cite{Stotyn:2013yka,Stotyn:2011ns}.
 
 The solutions of \cite{Herdeiro:2014goa}, \cite{Brihaye:2014nba} are constructed for a minimal potential
 of the scalar field: a mass term. Along the lines of \cite{Herdeiro:2015tia}, we study in this paper the
 influence of the non quadratic interaction on the solutions. The choice of the potential is explained
 in the paper. 
 
 Another feature of gravity in higher dimensions is the fact that the standard Einstein-Hilbert action
 can be supplemented by a hierarchy of terms involving higher powers of  the Riemann tensor. For $d=5$ there
 exist one such a term, the Gauss-Bonnet term and the general lagrangian for gravity  is know as 
 Einstein-Gauss-Bonnet (EGB) gravity.  
 The corresponding equations are in general more involved that the Einstein equation and the enhanced non linearity
 has generically an influence of the solutions.
 Five-dimensional spinning boson stars in EGB gravity were constructed in
 \cite{Henderson:2014dwa} with a negative cosmological constant and in \cite{Brihaye:2013zha,Brihaye:2015jja} with $\Lambda=0$.
 In both cases, it was found that the solutions exist for a limited range of the Gauss-Bonnet coupling constant. 
 Up to our knowledge, the existence of hairy black holes in higher dimensional EGB gravity was not emphasized yet.
 In this paper, we study these equations and we present strong  
 numerical evidence that the hairy black holes occurring in Einstein gravity can be extended to EGB gravity 
  up to a maximal value (depending of the angular velocity) of Gauss-Bonnet coupling constant.
 

The paper is organized as follows. 
In Section 2 we present the model, the ansatz and discuss some physical quantities which are relevant
for the understanding of the pattern of the solutions.
In Section 3 we describe the influence of an a self-interacting potential  on the Q-balls, the boson star
and the black holes. Section 4 is devoted to the construction of 
hairy black holes in Einstein-Gauss-Bonnet gravity. 
 Finally in section 5 we finish our discussion giving some further remarks.
\section{The Model}
We want to study spinning   boson stars and black holes
in 5-dimensional Einstein-Gauss-Bonnet (EGB) gravity.
The model is described by the following Lagrangian density
\be
\label{egbbs}
   S = \frac{1}{16 \pi G} \int d^5 x \sqrt{- g} ( R - 2 \Lambda + \frac{\alpha}{2} L_{GB} 
   - (16 \pi G) ( \partial_M \Pi^{\dagger} \partial^M \Pi + M_0^2 \Pi^{\dagger} \Pi + V_{si}(\Pi^{\dagger} \Pi ) )             
\ee
Here $R$ represents the Ricci scalar, $\Lambda=-6/\ell^2$ is the cosmological constant,
and $L_{GB}$ is the Gauss-Bonnet term. It is constructed out of the Riemann tensor in the standard way~:
\be
        L_{GB} = R^{MNKL} R_{MNKL} - 4 R^{MN} R_{MN} + R^2 \ \ , 
\ee
with $M,N,K,L \in \{0,1,2,3,4 \}$ and $\alpha$ denotes the Gauss-Bonnet coupling constant.   

The matter sector of the model consists of a doublet of complex scalar fields with the same mass $M_0$
and denoted by $\Pi$ in (\ref{egbbs}) where  we  supplemented a  self-interaction $V_{si}$. 
The full model presents an $U(2)$ global symmetry. 
The Noether current associated to the $U(1)$ subgroup,
i.e. $j^{A} = -i(\Pi^{\dagger} (\partial^{A}\Pi) -  (\partial^{A}\Pi^{\dagger}) \Pi))$, and the corresponding
conserved charge $Q$ will play an important role in the discussion of the solutions.
The variation of the action (\ref{egbbs}) with respect to the metric and the scalar field leads to the
Einstein-Gauss-Bonnet-Klein-Gordon equations. They are written in many places 
(namely in \cite{Brihaye:2014nba} with the same notations) and we do not repeat them here.

The Lagrangian (\ref{egbbs}) constitutes one of the simplest way to couple matter minimally to gravity.
Assuming $\Pi \neq 0$, the coupled system admits regular, localized, stationary solutions: the  boson stars 
(see e.g. \cite{Mielke:1997re} for a review). 
Boson stars in  $d > 4$ dimensional space-times were investigated namely in \cite{Astefanesei:2003qy,Prikas:2004fx,Hartmann:2013tca}.

\subsection{The ansatz}
In principle, spinning solutions in 5-dimensional space-time
can possess two independent angular momenta associated to the two orthogonal planes of rotation present in 4-dimensional
space. One  can however restrict oneself to the case of equal angular momenta.
Boson stars  possessing two equal angular momentum can be constructed in the above theory,
by using an appropriate anzatz for the metric and the scalar field \cite{Hartmann:2010pm}.
Interestingly, this ansatz leads to a system of differential equations.
Spinning boson stars in EGB gravity have been studied in details in \cite{Henderson:2014dwa} 
in the case $\Lambda < 0$ and  $M_0=0$.
Asymptotically flat, spinning boson stars and black holes  (with $M_0 > 0$) were  constructed  in 
\cite{Brihaye:2014nba} for Einstein gravity.  
The goal of  this paper is to study the influence of a self-interaction
and/or of the Gauss-Bonnet term on these  solutions. 
From now on, we will deal with  asymptotically flat solutions only.

{\bf Metric ansatz for the boson stars : }
 The relevant Ansatz for the metric reads 
\begin{eqnarray}
\label{metric}
ds^2 & = & -b(r) dt^2 + \frac{1}{f(r)} dr^2 + g(r) d\theta^2 + h(r)\sin^2\theta \left(d\varphi_1 - 
W(r) dt\right)^2 + h(r) \cos^2\theta\left(d\varphi_2 -W(r)dt\right)^2 \nonumber \\
&+& 
\left(g(r)-h(r)\right) \sin^2\theta \cos^2\theta (d\varphi_1 - d\varphi_2)^2 \ ,
\end{eqnarray}
where $\theta$ runs from $0$ to $\pi/2$, while $\varphi_1$ and $\varphi_2$ are 
in the range $[0,2\pi]$.
The corresponding space-time  possess two rotation planes at $\theta=0$ and $\theta=\pi/2$ 
and the natural $\mathbb{R}\times U(1)\times U(1)$ symmetry group is enhanced to $\mathbb{R}\times U(2)$. 
The metric above still leaves the diffeomorphisms related to the definitions of the radial variable $r$ unfixed. 
For the numerical construction of the solutions it is convenient to fix this  freedom by setting $g(r)=r^2$.

The key point of the construction is to choose the scalar doublet of the form 
\be
\label{harmonic}
          \Pi(x) = \phi(r) e^{i \omega t} \hat \Pi
\ee
where $\hat \Pi$ is a doublet of unit length that depends on the angular coordinates.
The standard non spinning solutions  are recovered by means of the particular form 
\begin{equation}
\label{hatphi_non}
 \hat \Pi = (1,0)^t  \ ,
\end{equation}
while for
rotating solutions the  parametrization is \cite{Hartmann:2010pm}
\be
\label{hatphi_rot}
\hat \Pi = \left(\sin \theta e^{i \varphi_1},\cos \theta e^{i \varphi_2}  \right)^t \ .
\ee  
This parametrization transforms the Einstein-Gauss-Bonnet-Klein-Gordon equations into a consistent set of differential equations.
 While the metric (\ref{metric})
has three commuting Killing vector fields $\partial_t$, $\partial_{\varphi_1}$, $\partial_{\varphi_2}$, 
the scalar doublet of the rotating solution with (\ref{hatphi_rot}) is invariant only 
under one possible combination of these vectors \cite{Dias:2011at}, namely 
\begin{equation}
 \partial_t - \omega\left(\partial_{\varphi_1} + \partial_{\varphi_2}\right) \ .
\end{equation}
The boson stars therefore have a single Killing vector.

{\bf Metric ansatz for the black holes : }
In order to construct the black holes, we used a slightly different ansatz which was first proposed in  
\cite{Brihaye:2014nba}
\begin{eqnarray}
\label{metric_bh}
ds^2 & = & -e^{2 F_0} N(r) dt^2 + e^{2 F_1} (\frac{dr^2}{N(r)} +   + r^2 d\theta^2)
+ e^{2 F_2} r^2 
 [\sin^2\theta \left(d\varphi_1 - W(r) dt\right)^2 +  \cos^2\theta\left(d\varphi_2 -W(r)dt\right)^2] \nonumber \\
&+& 
\left(e^{2F_1}-e^{2F_2}\right) r^2 \sin^2\theta \cos^2\theta (d\varphi_1 - d\varphi_2)^2 \ , \ N(r) = 1 - \frac{r_H^2}{r^2} \ ,
\end{eqnarray}
where $F_0,F_1,F_2$ are functions of $r$ only and $r_H$ denotes the position of the horizon. Note that
the radial variable $r$ has  different meaning in (\ref{metric}) and (\ref{metric_bh}). 
The above statement about the Killing symmetry still holds and
the spinning hairy black holes that we obtained possess a single Killing vector field.

The goal of this paper is to study the response of pattern of the 'minimal' boson stars and hairy black 
holes (i.e. in Einstein gravity and with a mass potential) to a self-interaction of the scalar field 
and  to the Gauss-Bonnet interaction. 

\subsection{Physical quantities}
The solutions can be characterized by several physical parameters, some of them are associated with the globally 
conserved quantities. The charge $Q$ associated to the U(1) symmetry has some relevance;
in terms of the ansatzs (\ref{metric}),(\ref{metric_bh}), the charge $Q$  takes respectively the forms
\begin{eqnarray} 
  Q = - \int \sqrt{- g} j^0 d^4 x &=& 4 \pi^2 \int_0^{\infty} \sqrt{\frac{bh}{f}} \frac{r^2}{b} (\omega + W) \phi^2 d r \nonumber \\ ,
  &=& 4 \pi^2 \int_{r_H}^{\infty} e^{-F_0 + 3  F_1 + F_2} \frac{ (\omega + W) }{N } \phi^2 d r
\end{eqnarray}
Interestingly, it was shown \cite{Hartmann:2010pm} that the charge $Q$ is related to the sum of the two (equal) angular
momentum of the solution~: $Q = 2 |J|$. The mass of the solution $M$
and the angular momentum $J$ can be obtained from the asymptotic decay of some
components of the metric~:
\begin{equation} 
                      g_{tt} = -1 + \frac{8 G M}{3 \pi  r^2} + o(\frac{1}{r^3}) \ , \
                      g_{\varphi_1 t} = - \frac{4 G J}{\pi r^2} \sin^2 \theta + o(\frac{1}{r^3}) \ , \
                      g_{\varphi_2 t} = - \frac{4 G J}{\pi r^2} \cos^2 \theta + o(\frac{1}{r^3})
\end{equation}
On the other hand, the black holes have an entropy and a temperature~:
\be
      A_H = 2 \pi^2 r_H^3 e^{2 F_1 + F_2}|_{r=r_H} \ \ , \ \ T_H = \frac{e^{F_0-F_1}}{2 \pi r}|_{r = r_H} \ .  
\ee
The speed of rotation of the black hole at the horizon is given by $W(r_H)$.
 
\subsection{Asymptotic and boundary conditions}
\subsubsection{Boson stars} The ansatz above transforms the field equations into a set of five differential equations in the five radial fields.
For boson stars, the equation for $f$ is of the first order while the equations for $b,h,W,\phi$ are of the second
order. 
The Taylor expansion of the fields about the origin takes the form
\be
  f = 1 + F_2 r^2 + o(r^4) \ , \ b = B_0 + B_2 r^2 + o(r^4) \ , \ h = r^2(1 + H_2 r^2 + o(r^4)) \ , 
\ee
\be  
  \ W = W_0 + W_2 r^2 + o(r^4) \ , \ \Pi = \Pi_1 r + \Pi_3 r^3 + o(r^5) \ ,
\ee 
where $F_2,B_0,B_2, \dots$ are constants.
The constants  $B_2$, $W_2$, $\Pi_3$ are determined in a straightforward way leading to 
\be
 B_2 = B_0  \frac{\alpha \Pi_1^2 + 3 (F_2 + H_2) - 6 / \ell^2}{3 \alpha(F_2+H_2)-3}  \ , \  W_2 = - \frac{\kappa \Pi_1^2 (W_0 + \omega)}{6 \alpha(F_2 + 3 G_2)- 6}
\ee
and a lengthy expression for $\Pi_3$. The  constants $ F_2, H_2$ are related by the condition 
 \be
\label{constraint}
   3 \alpha( F_2^2 + 2 F_2 H_2 + 5 H_2^2) - 6(F_2 + H_2) - 2 \kappa \Pi_1^2  = 0 \ \ .
\ee
which is linear if $\alpha = 0$ and quadratic for EGB gravity $\alpha >0$.
In principle,  all the orders of the Taylor expansion are determined in terms of $B_0,  W_0,  \Pi_1$ and  $H_2$ (or alternatively of $F_2$ obeying (\ref{constraint})). For $\alpha >0$, the constraint imposes some limit
in $\alpha$ for spinning boson stars of definite angular momentum \cite{Brihaye:2015jja}.

In the asymptotic region, the metric has to approach the Minkowski space-time and the scalar field vanishes.
Finally, the single parameter $\omega$ is sufficient to characterize a solution.
\subsubsection{Black holes}
The ansatz adopted for black holes leads to five second order equations. The regularity at the horizon
imposes the {\it derivatives} of the  fields $F_j, W$ ($j=0,1,2$) to vanish at the horizon and $W(r_H) = \omega$.
So, the angular velocity on the horizon of the black holes should coincide with the harmonic frequency $\omega$
of the scalar field set in (\ref{harmonic}). 
As for boson stars, all fields should vanish in the limit $r \to \infty$. 
The parameters $r_H$ and $\omega$ are, in principle, sufficient to characterize 
a black hole. We will see that the domain of existence is quite limited
and that -on this domain-  more than one solution can correspond to a choice $r_H,\omega$.

\section{Self-Interacting solutions in Einstein gravity}
Here we want to construct the bosons star and Q-ball associated to a polynomial potentials of the form
\be
        V_{si} = \lambda_6 |\Phi|^6 - \lambda_4 |\Phi|^4  \ .
\ee
Because of the numerous parameters, we will put the emphasis on two particular  cases
which can be considered as extremal:
\begin{itemize}
\item  The case $\lambda_6=0$, $\lambda_4=0$. This case then corresponds  
 to a mass term only and the corresponding solutions have been studied in details in 
\cite{Brihaye:2014nba}, in the case of Einstein gravity. 
This mass potential will be denoted $V_2$ in the following.
As demonstrated in \cite{Volkov:2002aj} the  boson stars  do not admit Q-balls counterpart in this case.
The gravitating parameter $16 \pi G$ can be rescaled in the scalar field.
The solutions corresponding to EGB gravity  will be presented for this potential only. 
\item The case $\lambda_4 = 2 M_0^2/\phi_0^2$,  $ \lambda_6  =  M_0^2/ \phi_0^4$. 
This choice corresponds to a positive
definite potential presenting two degenerate local minima at $\phi = 0$ and $\phi = \phi_0$. 
It was proposed \cite{Friedberg:1986tq} and allow for solutions to exist in the absence of gravity.
In lower dimensions, this potential was  used  for the study of kink collisions, see e.g. \cite{Dorey:2011yw}.
Four-dimensional charged boson stars have been constructed in  \cite{Brihaye:2015veu} with such a potential.
In the following the corresponding potential will be denoted $V_6$. 
For this potential, the gravity constant and the value $\phi_0$ can be combined into a dimensionless quantity
 noted  $\kappa \equiv 16 \pi G \phi_0^2$. 
\end{itemize}

The field equations are coupled, non-linear differential equations with boundary conditions.
To our knowledge, they do not admit closed form solutions
and approximation techniques have to be used. 
For instance, the  relevant solutions can be obtained numerically
by a fine tuning of the parameters $\omega$ and $\phi'(0)$ 
in such a way that the boundary conditions are obeyed.  
For our numerical construction we  used the numerical 
solver COLSYS \cite{colsys} based on the Newton-Rafson algorithm.

\subsection{Q-balls}
We first present the solutions obtained in the absence of gravity (i.e. with $\kappa = 0$). 
The only parameter to vary is the value $\phi'(0)$ (or equivalently the frequency  $\omega$). 
Along with boson stars in four dimensions, families of  solutions exist 
with different numbers of nodes of the scalar function $\phi(r)$.
For simplicity, we addressed only the solutions for which the scalar function has no nodes. The solutions with nodes
are usually considered as excited modes of these 'fundamental' one. 
 \begin{figure}[ht!]
\begin{center}
{\label{v_6}\includegraphics[width=8cm]{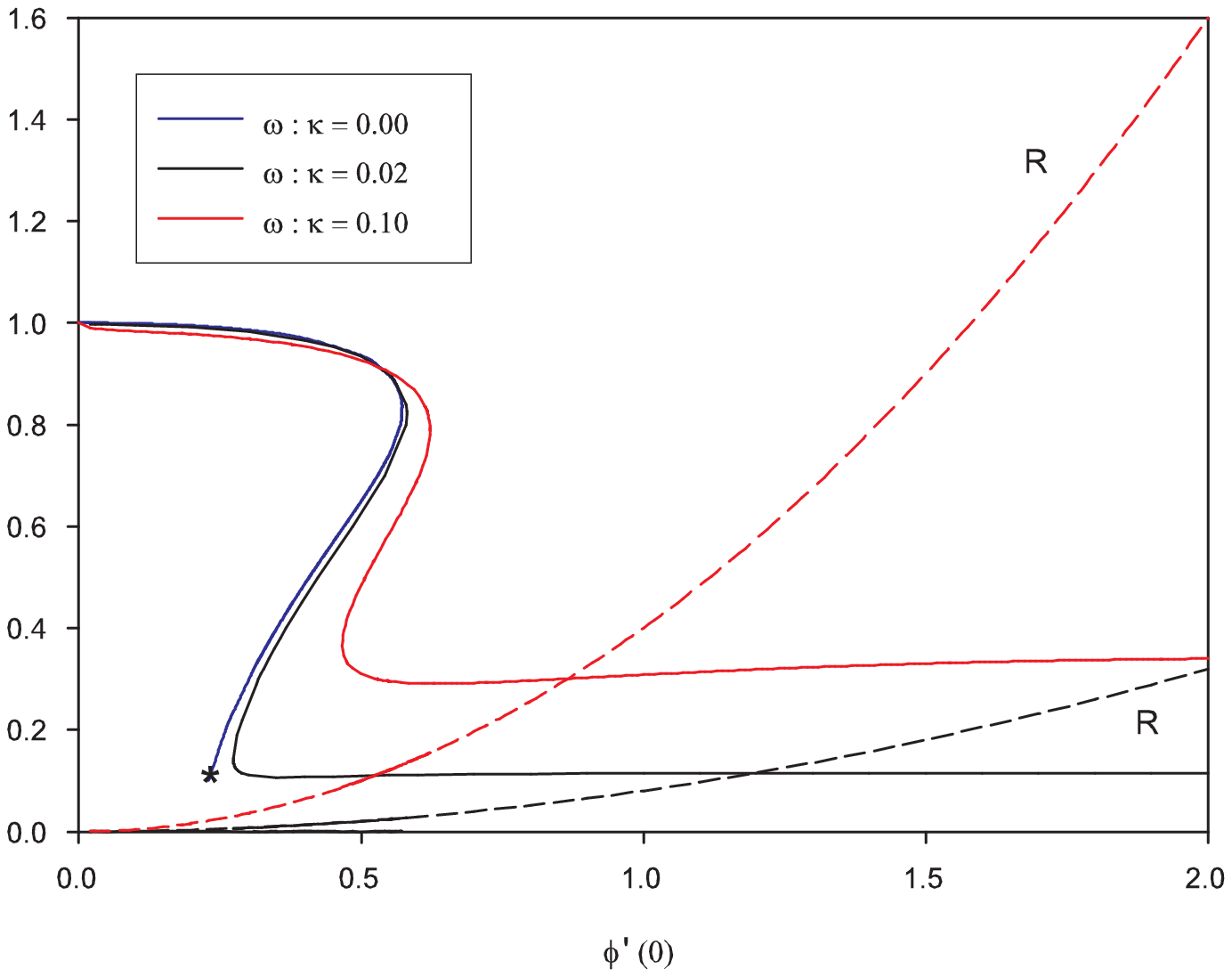}}
{\label{v_v_r}\includegraphics[width=8cm]{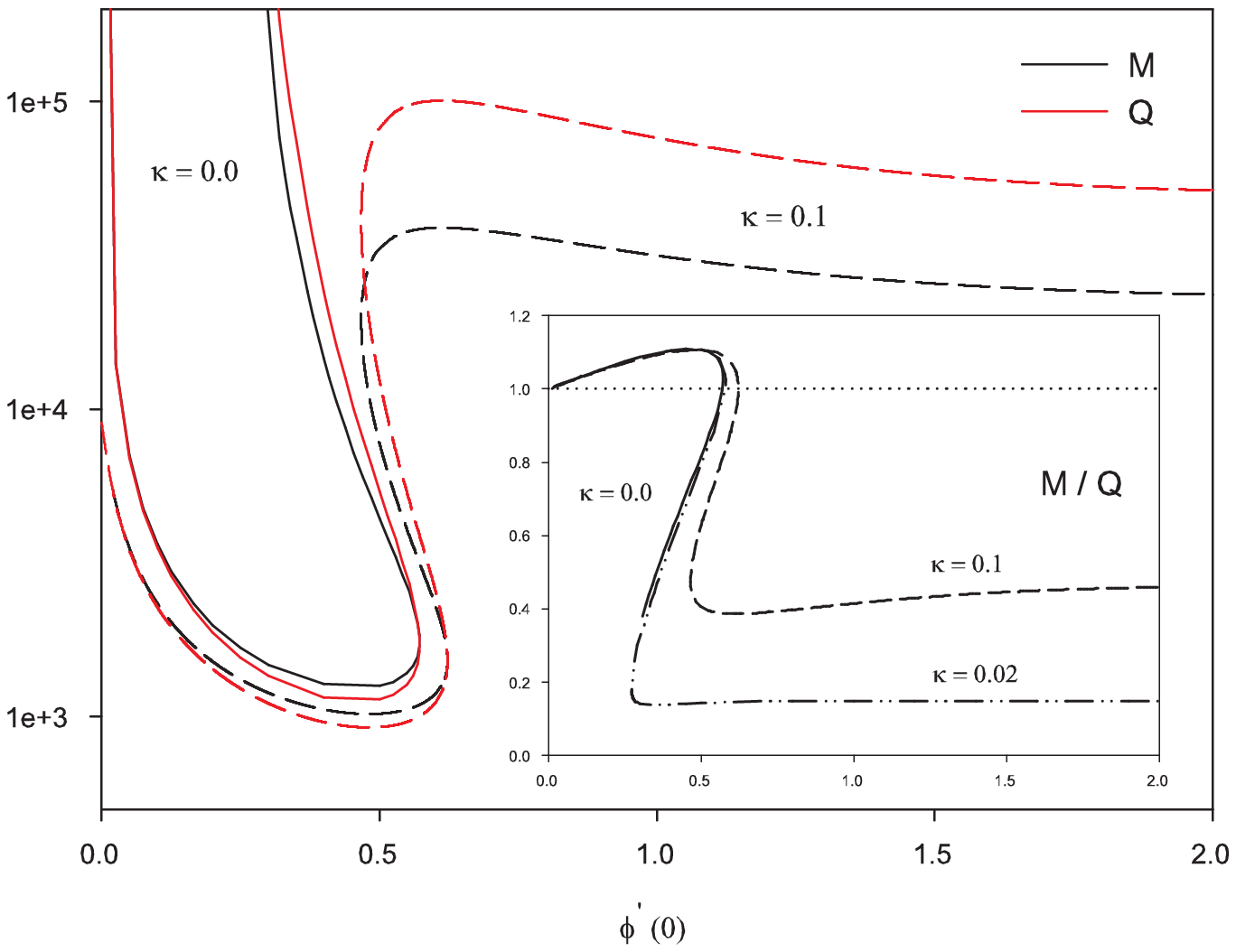}}
\end{center}
\caption{Left: The frequency $\omega$ and the value $R(0)$ as functions of $\phi'(0)$ for Q-balls ($\kappa = 0$)
and boson stars $(\kappa = 0.1 , \kappa = 0.02)$.
Right: Mass and  charge of the solutions. The ratio $M/Q$ is plotted in the insert.
\label{data_v_6}
}
\end{figure}
The Q-balls can be constructed numerically by increasing progressively the control parameter $\phi'(0)$,
forming a branch to which we refer as the main branch.
In the limit $\phi'(0) \to 0$, the vacuum configuration $\phi(r)=0$ is uniformly approached while $\omega \to 1$.
The convergence is slow and quantities like the mass (and also the charge)
converge to a finite value, forming a mass gap. As demonstrated by Fig. \ref{data_v_6}, the
solutions exist up to a maximal value of parameter $\phi'(0)$; we find $\phi'(0)_{max} \approx 0.572$
corresponding to a frequency $\omega \approx 0.84$. Solving the equations for lower values of the frequency reveals
the existence of another branch of Q-balls, let us call it the second branch.
We found strong indications that these solutions exist for $\phi'(0) \in ]\phi'(0)_{min},\phi'(0)_{max} ]$ and 
$\omega \in ]0, 0.84[$ (with $\phi'(0)_{min} \approx 0.25$).
The numerical construction of the solutions of the second branch 
is  difficult for the small values of $\omega$ (the {\it star '$\star$'} on Fig. \ref{data_v_6}
represents the limit where our numerical solutions became unreliable).  In this region,
the profile of the function $\phi(r)$ presents a pronounced abrupt wall for $r \approx R_w$
(we define the parameter  $R_w$  as the value of $r$
 where the function  $\phi'(r)$ attains its  local minimum).
This wall  separates two plateaus; the first one has $\phi(r) \sim 1$ (for $r < R_w$) and 
the second at $\phi = 0$ (for $r > R_w$). 
As an   illustration of this result  the profiles of the two solitons corresponding to $\phi'(0) = 0.3$ (one for each branch) 
are shown on Fig. \ref{q_ball_1}.  
The numerical results show that, in  the limit $\phi'(0) \to \phi'(0)_{min}$, 
the mass, the charge and the  radius $R_w$ of the solution tends to infinity: 
this can be related to the fact that no attractive force (like gravity)
is operating, as a consequence the star can spread arbitrarily large in space. 
\begin{figure}[ht!]
\begin{center}
{\label{vv_6}\includegraphics[width=10cm]{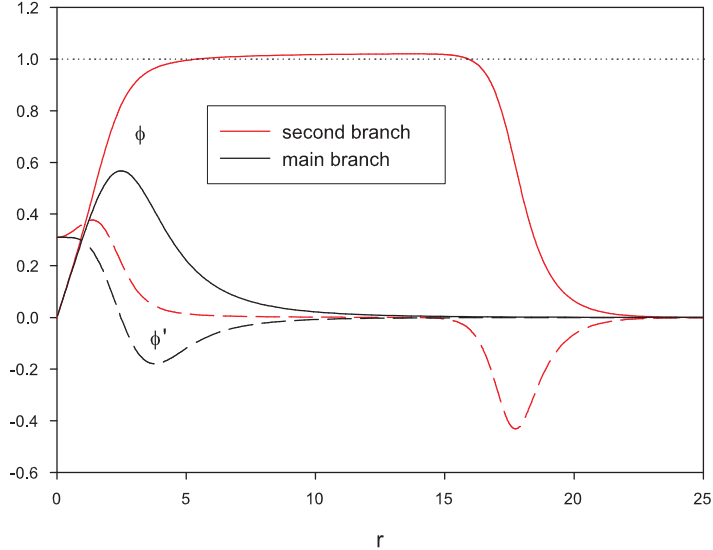}}
\end{center}
\caption{ The profiles $\phi$ and $\phi'$ for the two Q-balls (one on each branch) corresponding to $\phi'(0) = 0.3$.
\label{q_ball_1}
}
\end{figure}
The solutions on the main  branch correspond to high frequencies
$ 1 > \omega > 0.84$ and have $M/Q > 1$;
 interestingly the solutions of  second branch correspond to low frequencies
($ 0.84 > \omega > 0$) and  have
 $M/Q < 1$.  Following standard arguments about the classical stability of Q-balls, only 
 the solutions of the second branch  are classically stable.
\subsection{Boson stars}
Boson stars correspond to deformations of the Q-balls (see previous section) for $\kappa > 0$.
Once coupled to gravity the pattern of the solutions change qualitatively. 
The major difference between boson stars
and Q-balls is that -contrary to Q-balls- boson stars exist for large values of $\phi'(0)$.
Several features of Q-balls and bosons stars are illustrated on Fig. \ref{data_v_6} for three values of $\kappa$.  

For fixed $\kappa$ and $\phi'(0)$ increasing, the value $b(0)$ approaches zero while 
the value  of the Ricci scalar at the origin (say $R \equiv R(0)$) 
  increases  roughly quadratically. 
This suggests that a singular configuration is approached 
in the limit $\phi'(0) \to \infty$. Outside the central region (we find typically for $r \in [R_w/10, \infty]$)
the metric and the  the scalar field are rather intensive to the increase of $\phi'(0)$. In particular, $\omega, M,Q, R_w$
approaches constants for $\phi'(0) \to \infty$ (these constants depend, off course, on $\kappa$). 
Note that no counterpart of the second branch of Q-ball is found. This phenomenon can be explained by the fact that
the attracting gravitational interaction keeps the matter field relatively compact
and does not allow for an  extension in an arbitrarily large volume.

Since Q-balls ($\kappa = 0$) do not exist for $\phi'(0) > \phi'(0)_{max}$, it is  natural
to examine   the behavior of boson stars ($\kappa > 0$) for fixed values of $\phi'(0)$, greater than $\phi'(0)_{max}$, and decreasing the parameter $\kappa$.
It turns out that the parameters $R_w$ (and similarly $M,Q$) increases, likely tending to infinity
in the limit $\kappa \to 0$.
We find in particular $R_w \sim 1/\kappa$: as it could be expected,
the decreases gravity, through the effective coupling parameter $\kappa$, allows for compact object with
larger and larger size 
to form. 
\begin{figure}[ht!]
\begin{center}
\subfigure[Matter field]
{\label{vvv_6}\includegraphics[width=8cm]{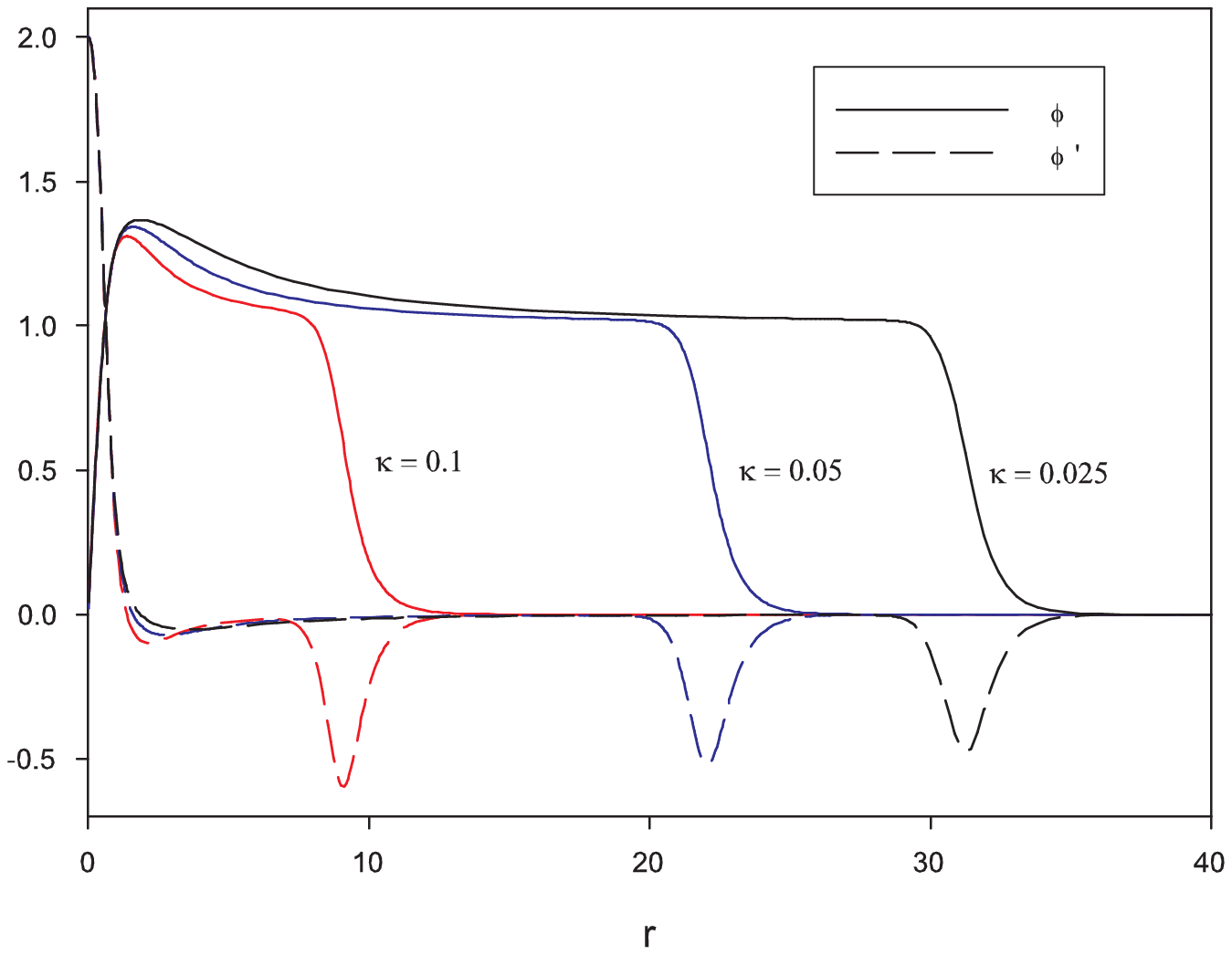}}
\subfigure[Metric function]
{\label{vvv_v_r}\includegraphics[width=8cm]{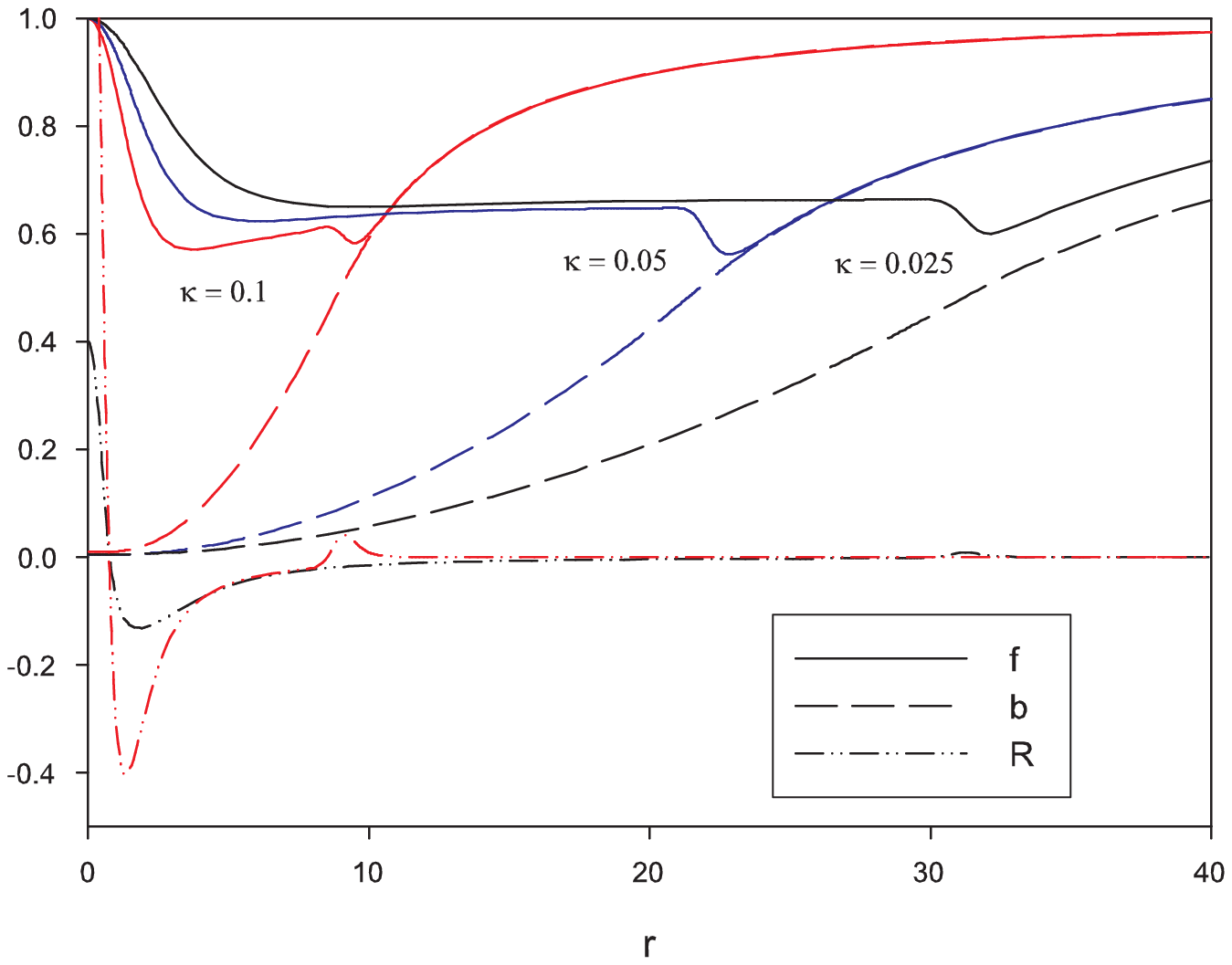}}
\end{center}
\caption{Profiles of the scalar and metric functions for $\kappa = 0.1, 0.05$ and $\kappa = 0.025$.
\label{profile_al_to_0}
}
\end{figure}
To illustrate this statement, the profiles of the different fields corresponding
to $\kappa = 0.1, 0.05$ and $\kappa = 0.025$ are presented on Fig. \ref{profile_al_to_0}.
We further notice  that the Ricci scalar $R(r)$  approaches uniformly the null function although the metric
potentials keep  non trivial profiles.

\begin{figure}[ht!]
\begin{center}
{\label{vvvv_6}\includegraphics[width=10cm]{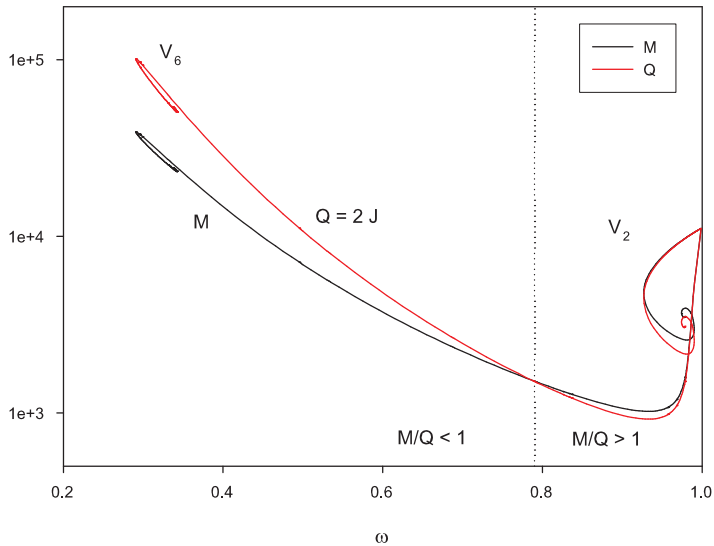}}
\end{center}
\caption{Mass and charge of minimal and self-interacting boson star corresponding to $\kappa = 0.1$.
\label{fig_1}
}
\end{figure}

{\bf Comparison of the $V_2$ and $V_6$ potentials : }
As pointed out already, boson stars exist with the  mass potential $V_2$. On Fig. \ref{fig_1} the charge and the
mass of the minimal and self-interacting boson stars are superposed. The non linear interaction clearly allows for
solutions with a lower frequency, then also corresponding to a slower speed of rotation on the horizon. 
The minimal boson stars have $M/Q > 1$ on their whole domain
of $w$. Accordingly, they have no positive binging energy and are classically unstable. 
The interaction makes the solutions  classically stable for the low values of the frequency. 
\subsection{Black holes}
The construction of black holes in Einstein gravity and for the mass potential constitutes the object of \cite{Brihaye:2014nba}.
It was found that black holes exist in a very specific domain of the parameters $w,r_H$  
(see in particular Fig. 1 of \cite{Brihaye:2014nba}). 
Here we will present the influence of the self-interaction $V_6$ on the families of black holes obtained in that paper.

The pattern of solutions is quite involved. It turns out that,
corresponding to a fixed value of $w$, two (and sometimes even four) branches of black holes exist 
on  specific intervals of the horizon parameter $r_H$. 
In order to describe the pattern,
let us choose one particular boson star characterized by a frequency $w$.
 Integrating the black holes equations for this value of $w$  we find that one branch of black holes exist
for   $r_H \in ]0,r_{H,max}]$. In the limit $r_H \to 0$  the limiting solution on this branch approaches the 
starting boson star; correspondingly, the temperature $T_H$ diverges  while the horizon area $A_H$ tends to zero.
In the following, we will refer to such a family of black holes as to the {\it BS-branch}.
The data corresponding to $w = 0.995$ (i.e. close to the limiting value $w=1$) is reported on Fig. \ref{data_w_0995}.
\begin{figure}[ht!]
\begin{center}
{\label{w_6}\includegraphics[width=8cm]{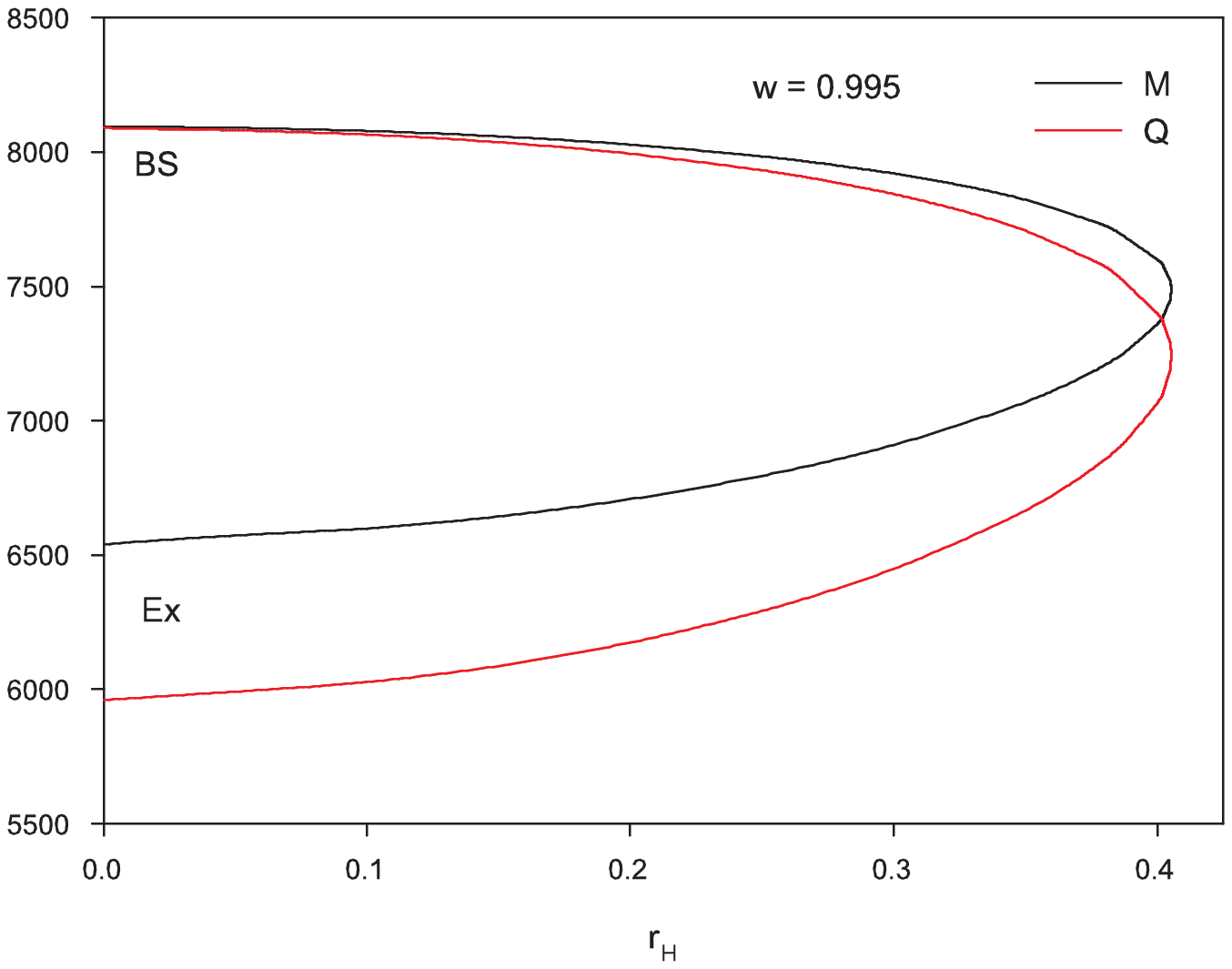}}
{\label{w_w_r}\includegraphics[width=8cm]{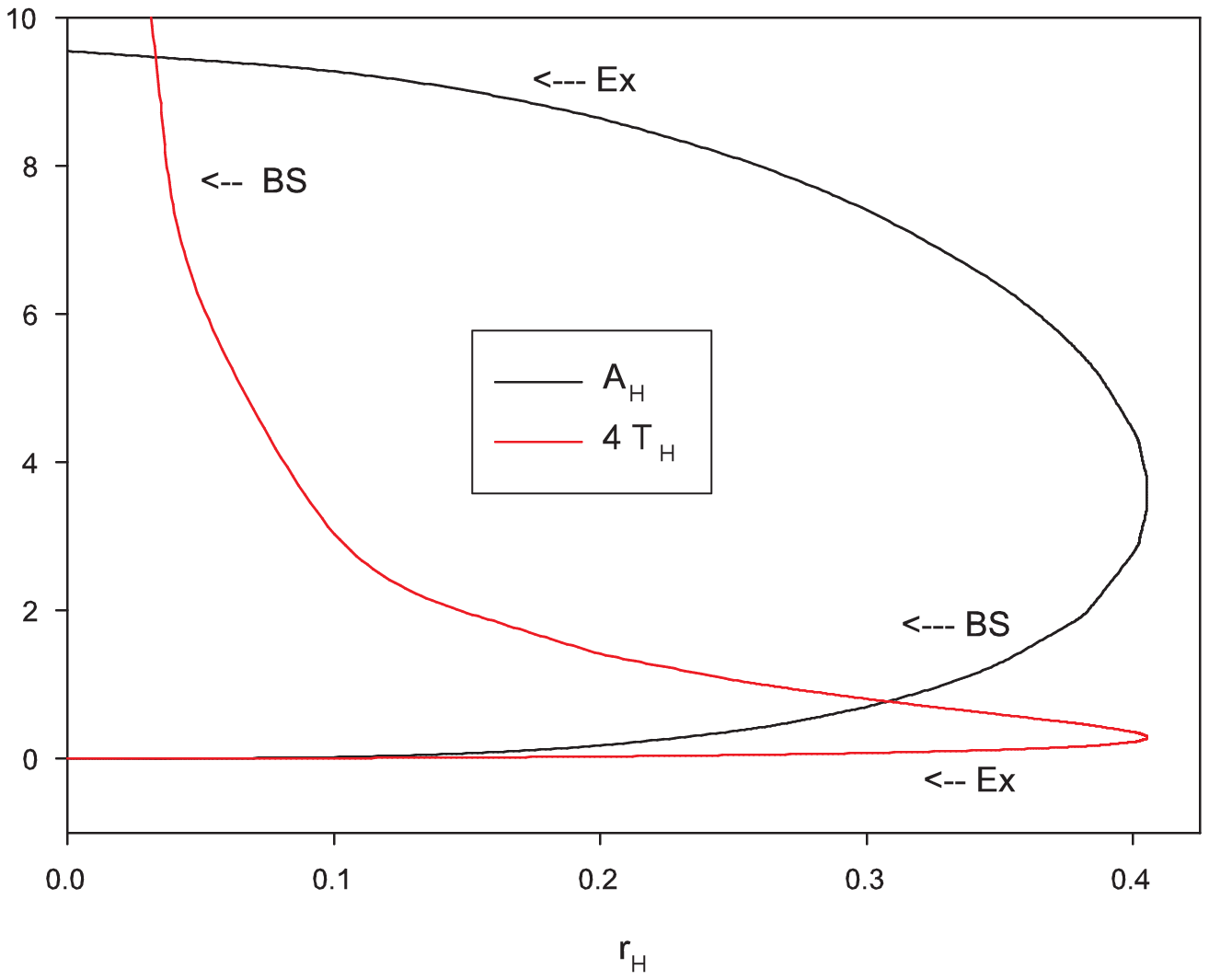}}
\end{center}
\caption{
Left: Mass and  charge of the black hole for $w = 0.995$ and $\kappa = 0.1$ as function of the horizon $r_H$.
Right: Temperature and horizon area as functions of $r_H$.
\label{data_w_0995}
}
\end{figure}
A careful study of the equations for the {\bf same} value of $w$  reveal that another branch of solutions exists,
also labeled by the  horizon parameter $r_H$.
In the limit $r_H \to 0$ the solutions of this second branch  terminate either into an extremal black hole 
(in this case the temperature $T_H$ tends to zero) or into another boson star with the same $w$.
Setting $w=0.995$ (see Fig. \ref{data_w_0995}) the second branch indeed terminates into an extremal black hole;
we labeled it the $Ex$-branch. The $Ex$-branch and $BS$-branch coincide for $r_H = r_{H,max}$.
We found that the value $r_{H,max}$ decreases  slowly with $w$; for example $r_{H,max} \approx 0.4$ 
for $w = 0.995$ and  $r_{H,max} \approx 0.36$ for $w = 0.3$. 
\begin{figure}[ht!]
\begin{center}
{\label{ww_6}\includegraphics[width=8cm]{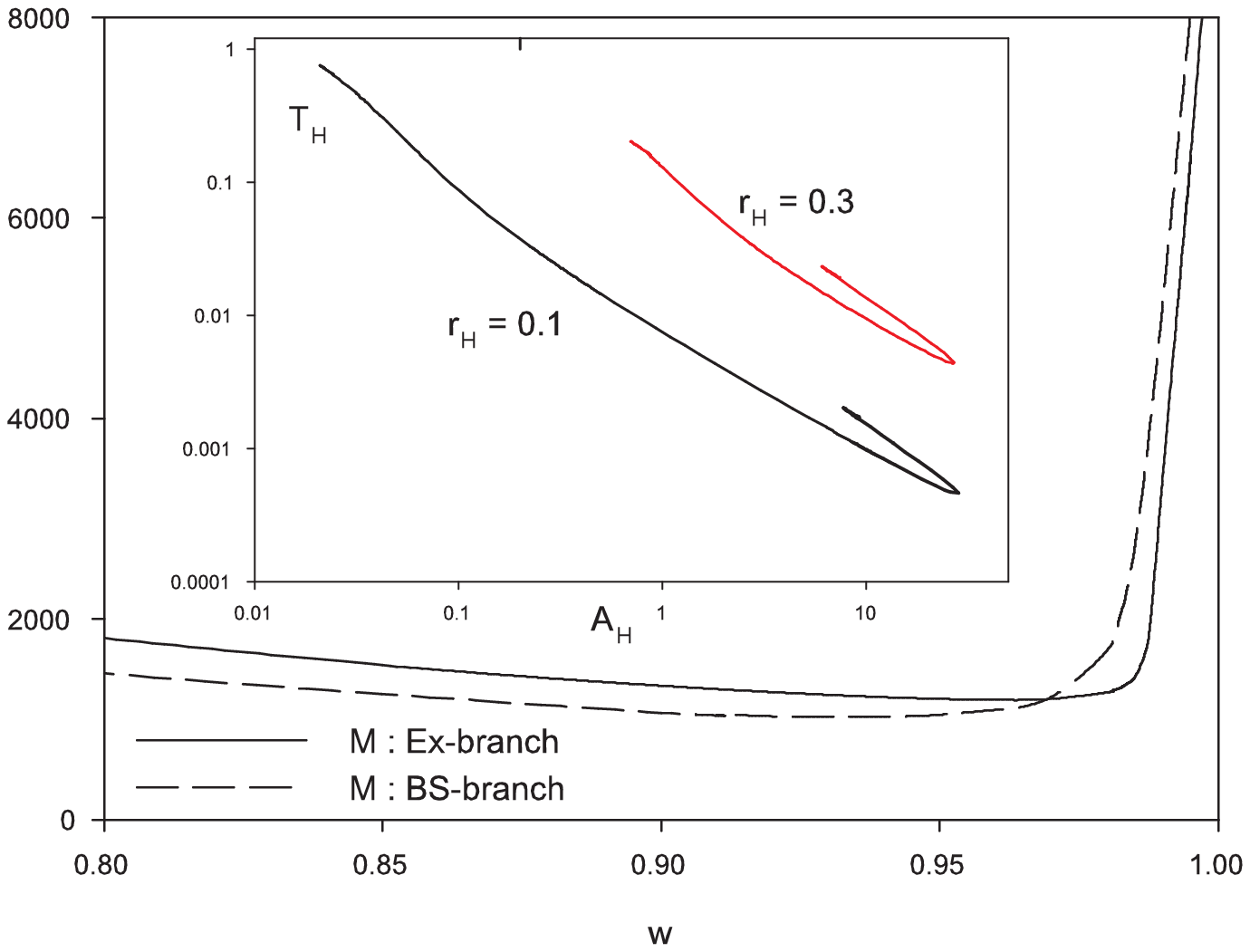}}
{\label{ww_w_r}\includegraphics[width=8cm]{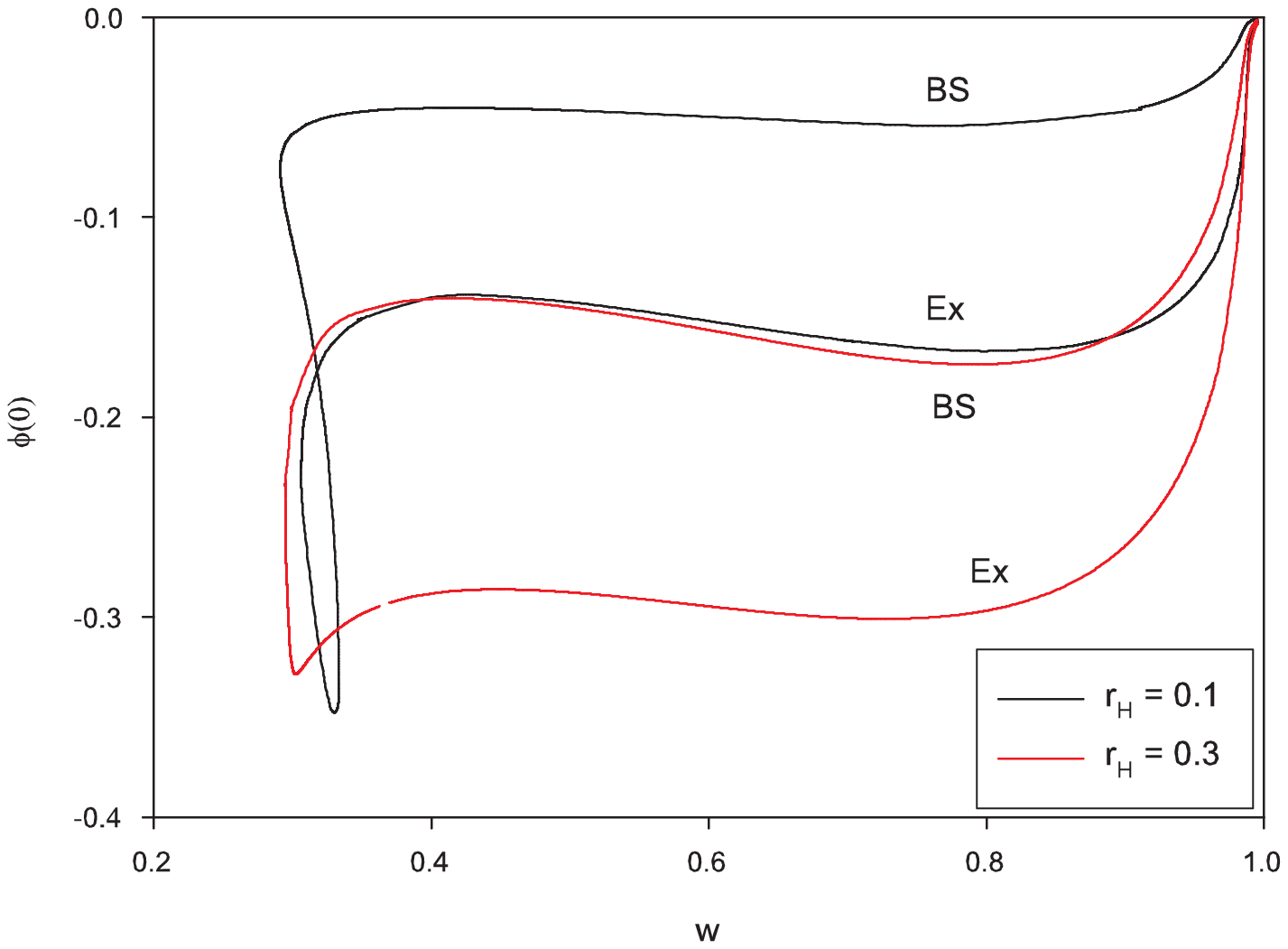}}
\end{center}
\caption{
Left:
Mass of the boson star (dashed line)  and of the extrema solution (solid line)
for $\kappa = 0.1$ as function $w$  region $w \sim 1$. 
Insert : Temperature as function of the horizon area for two values of $r_H$.
Right: 
Value of the scalar field at the horizon as function of $w$ for two values of $r_H$.
\label{data_xh_01_a}
}
\end{figure}
\begin{figure}[ht!]
\begin{center}
{\label{www_6}\includegraphics[width=8cm]{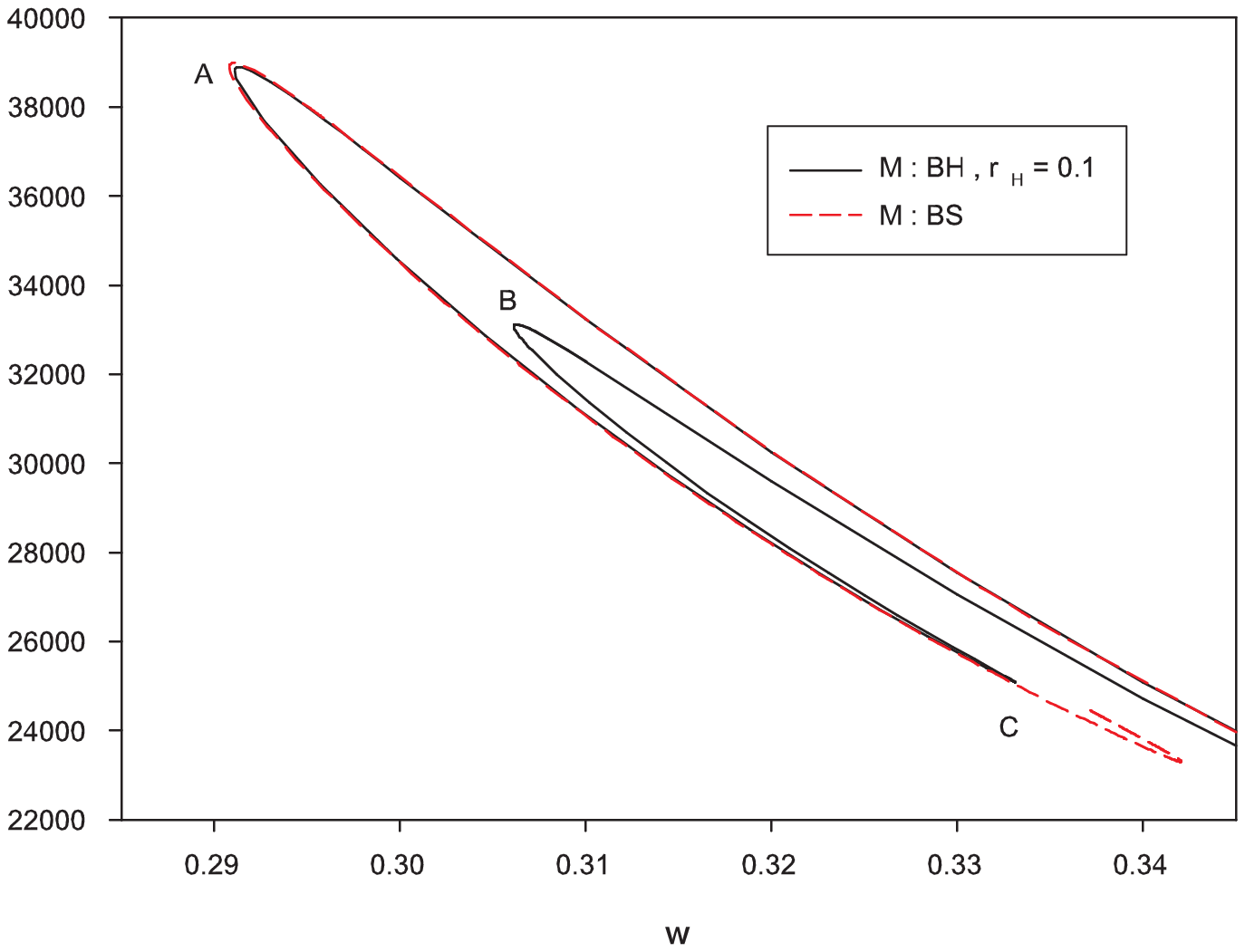}}
{\label{www_w_r}\includegraphics[width=8cm]{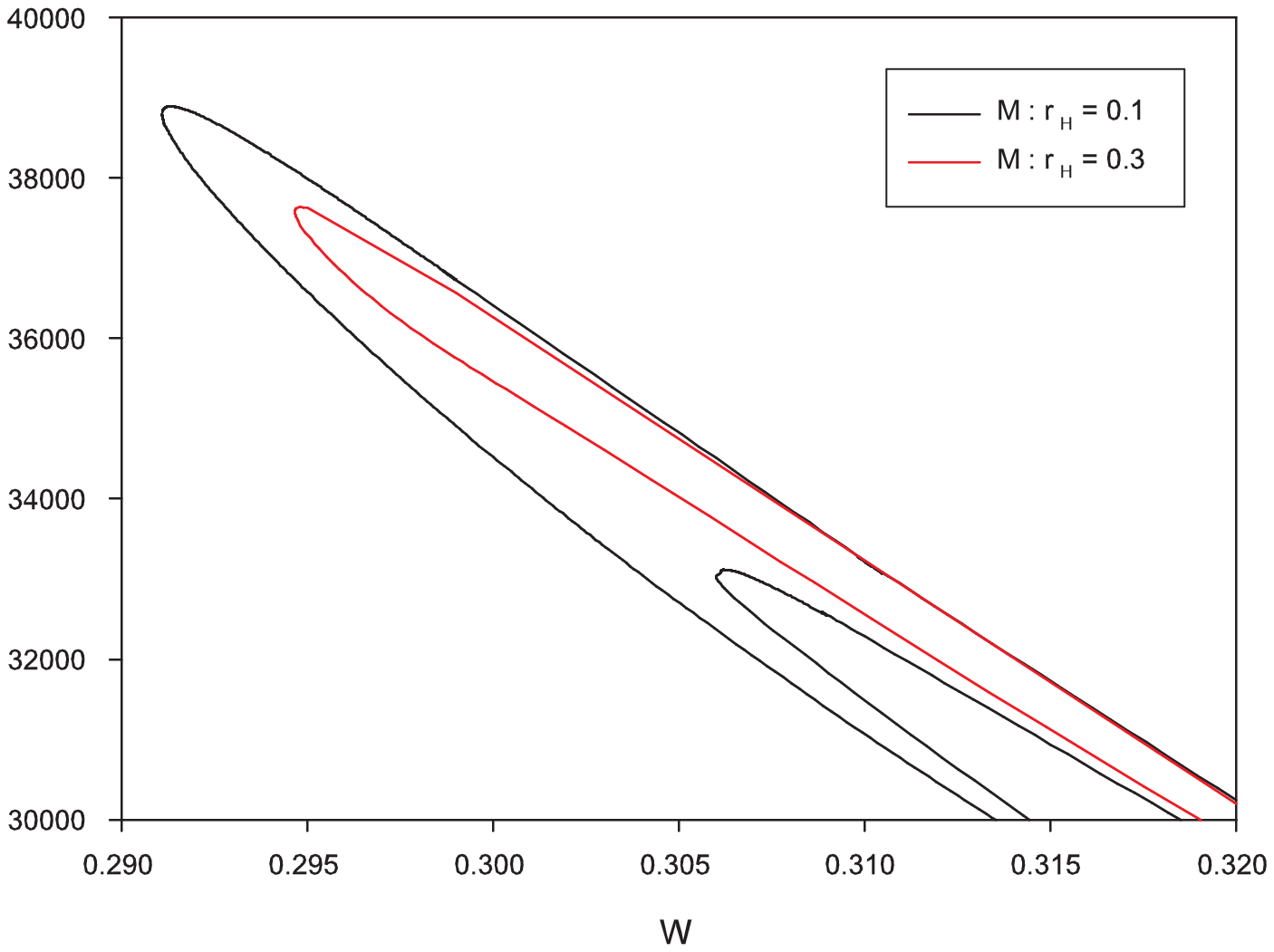}}
\end{center}
\caption{
Left:
Mass of the black hole (black line) with $r_H = 0.1$ and $\kappa = 0.1$ as function $w$ in the critical region;
The mass of the boson  star  (dashed red).
Right: 
Mass of the black hole with $r_H = 0.1$ and $r_H = 0.3$ (for  $\kappa = 0.1$) as function $w$ in the critical region. 
\label{data_b}
}
\end{figure}
Repeating this operation with different values of $w$  to determine the domain of existence
of the solutions is quite demanding; we therefore employed a slightly different strategy.  
 We constructed the families of black holes corresponding to fixed values of $r_H$ and varying $w$. 
Several features of the pattern  are illustrated 
 on Figs. \ref{data_xh_01_a},\ref{data_xh_01_b} where some parameters (mainly the mass) 
 are reported in function of $w$. 
 These figures, where for definiteness we set $\kappa = 0.1$, appeals several comments~:
\begin{itemize} 
\item On Fig.\ref{data_xh_01_a} (left part) the mass of
 boson star and of the  extremal solution are given  in the region $w \sim 1$.
\item The values of the scalar field on the horizon for the black holes with $r_H= 0.1$  and $r_H = 0.3$ 
are superposed on right side of Fig. \ref{data_xh_01_a}. The $A_H, T_H$ relation for these black holes
is supplemented in the insert in the left side of the figure.
\item For $w > 0.97$ and $w < 0.35$  we have  $M_{BS} > M_{Ex}$.
For intermediate values of $w$ we have instead $M_{BS} < M_{Ex}$.
This contrasts with the case of the mass term (see \cite{Brihaye:2014nba}).
\item  Figs. \ref{data_b} show the mass-frequency dependence in the region 
of the small frequencies. The left side shows a comparison between the mass of boson star
(dashed red line)  and the mass of the black hole corresponding to $r_H = 0.1$ (black line).
 The curve corresponding to  the black holes  of the $BS$-branch  is all long 
very close to the curve corrsponding  to the boson star.
These curves can hardly be distinguished on the graphic but 
separate only when the boson star mass starts spiraling. 
For each $w$ such that  $w \in [0.307, 0.334]$ four black holes  with different masses exist.  
The $BS$-branch and $Ex$-branch join at the point marked $C$ corresponding to $w\approx 0.334$ and $M \approx 25.000$. 
This value is very close to the mass of the corresponding boson star. 
\item Increasing the horizon value, the pattern get simpler. In particular the points marked $B,C$ 
on the figure disappear and only two solutions persist
for all accessible values of $w$. The masses corresponding to $r_H= 0.1$  and $r_H = 0.3$ 
are compared on Fig. \ref{data_b} (right part).
\end{itemize}

As Figs. \ref{data_a} and \ref{data_xh_01_b} reveal, the pattern of the black holes solutions is quite involved
and uneasy to describe solely with the  control parameters $r_H$ and $w$. On Fig. \ref{th_mass}
we illustrate another feature of the solutions by using a mass-temperature plot 
 for the solutions corresponding to $r_H = 0.1$  and $r_H = 0.3$. 
In each case, the points where the curves stop correspond to the limit $w \to 1$.
The use of $T_H$ as parameter allows one to disentangled the BS and Ex-branches. 
 The range of accessible
temperature for a definite $r_H$ decreases while $r_H$ increases and collapse for $r_H \sim 0.4$.
The three points labeled $A,B,C$ on this graphic correspond to the critical values marked also in Fig. {\ref{data_b}. 
\begin{figure}[ht!]
\begin{center}
{\label{wwww_6}\includegraphics[width=10cm]{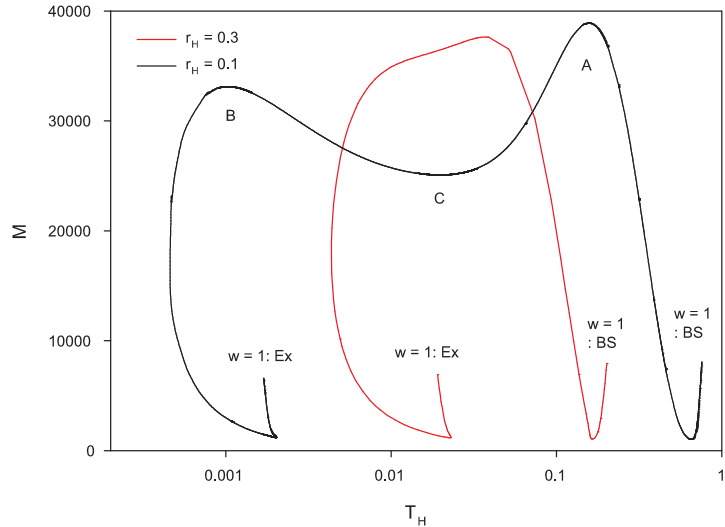}}
\end{center}
\caption{Mass-Temperature relation for black holes with $r_H = 0.1$ and $r_H = 0.3$ (here $\kappa = 0.1$).
\label{th_mass}
}
\end{figure}

To conclude this section let us stress a few differences between the families of black holes associated
with the $V_2$ and $V_6$ potentials~:
\begin{itemize}
	\item[(i)] The black holes can exist 
for smaller values of $w$ (i.e.  angular velocity at the horizon) than for the mass potential.
	\item[(ii)] Our results suggest that considering smaller values of $\kappa$
 would  result into the occurrence of black holes with arbitrarily small values of $w$.
 The numerical construction becomes however difficult.
	\item[(iii)] The black holes with the low values for $w$ have much 
higher masses (and charge $Q$) than the ones close to the limit $w = 1$.
\end{itemize}

\section{Solutions in EGB gravity}
As said already, the influence of the Gauss-Bonnet term on spinning boson stars existing in 5-dimensional gravity 
was studied by several authors. A general feature seems to be that any solution of this type
can be deformed continuously by progressively increasing the Gauss-Bonnet parameter $\alpha$ and 
ceases to  exist at a critical  value, say $\alpha_{max}$. 
\begin{figure}[ht!]
\begin{center}
{\label{x_6}\includegraphics[width=8cm]{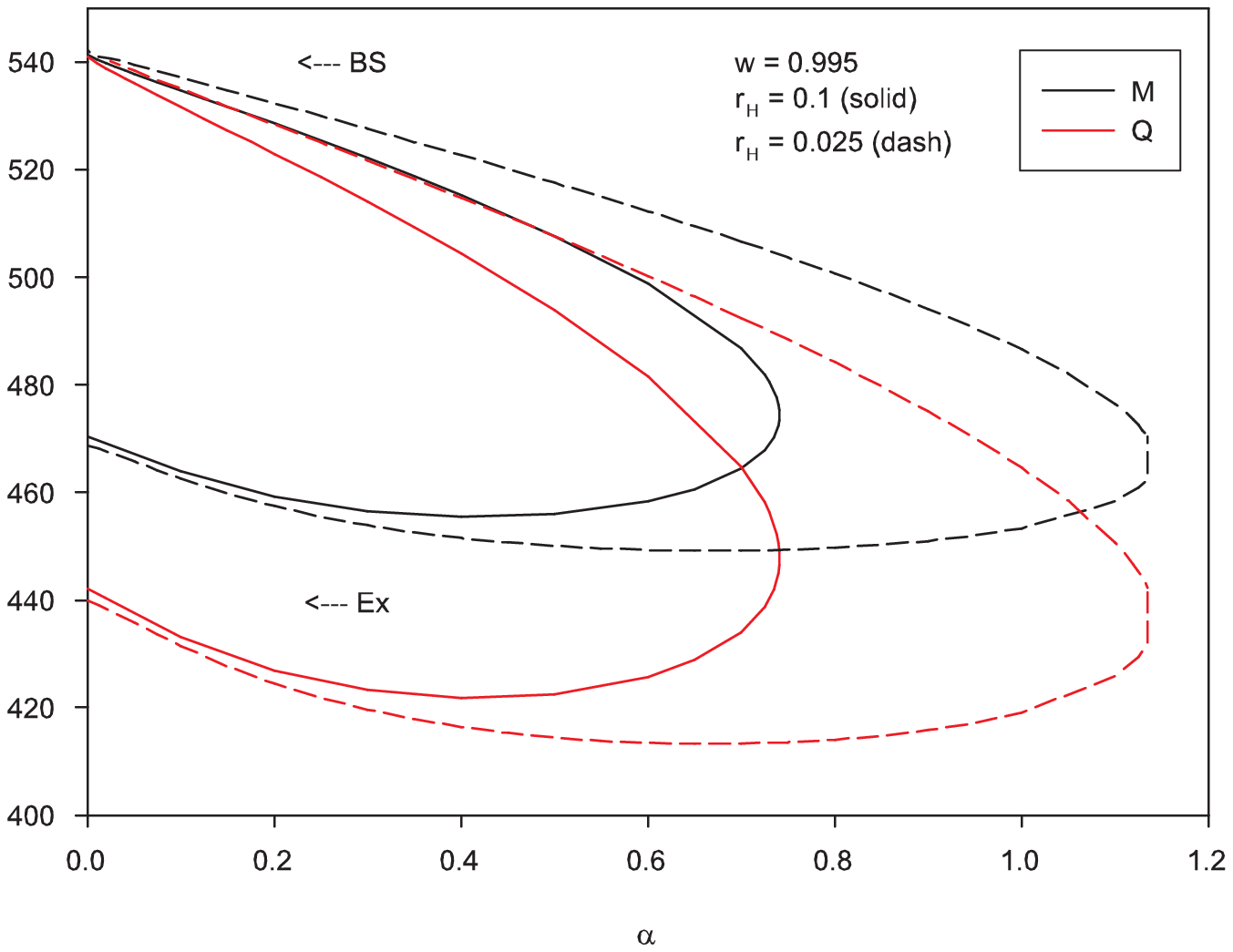}}
{\label{x_r}\includegraphics[width=8cm]{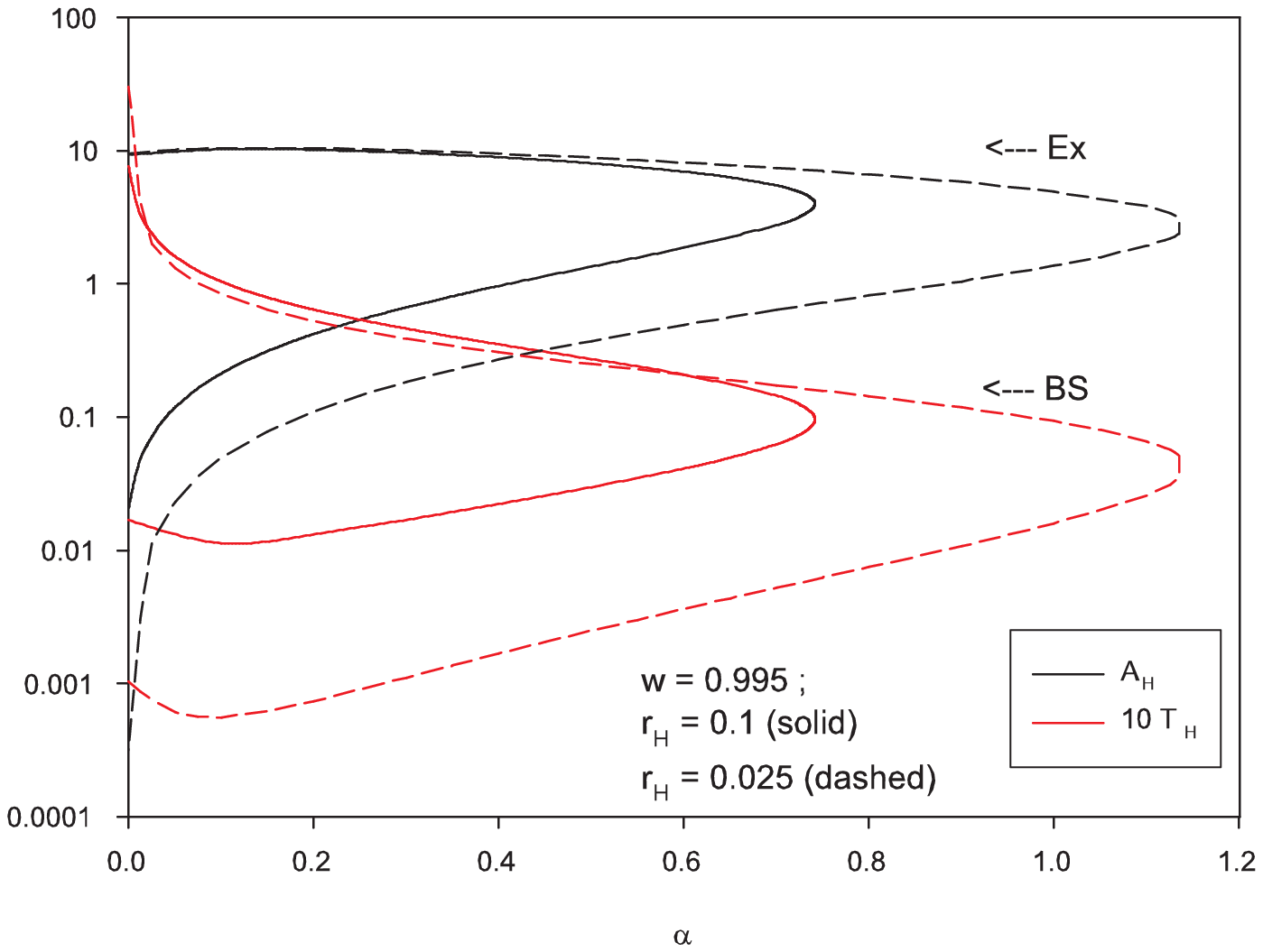}}
\end{center}
\caption{
Left:
Mass and charge of black holes with $w = 0.995$ and $r_H = 0.1$ (solid) and $r_H = 0.025$ (dashed)
as function $\alpha$. 
Right: 
Horizon area and temperature of black holes with $w = 0.995$ and $r_H = 0.1$ (solid) and $r_H = 0.025$ (dashed)
 as function $\alpha$.
\label{gb_w0995}
}
\end{figure}

The reasons of this limitation can be found analytically  \cite{Henderson:2014dwa}, \cite{Brihaye:2015jja}
by performing a Taylor expansion of the solutions around the origin. 
It turns out that Gauss-Bonnet interaction implies some 
quadratic constraints like (\ref{constraint}) between the Taylor coefficients; as a consequence 
 some coefficients might become complex for large values of $\alpha$ and the solution stops to be real.
Since the black holes are only considered outside their event horizon, the question of their existence
for $\alpha > 0$ raises naturally. 

\begin{table}[ht!]
\begin{center}
\caption{Critical value of $\alpha$ for a few values of $w$ and $r_H$.
The two first lines correspond to the branch with the higher mass and the last line corresponds to the branch with the lower mass.}
\begin{tabular}{c|c c}
$\alpha_{max}$ & $r_H$ = 0.025 & $r_H$ = 0.1 \\ \hline $w$ = 0.995 & 1.15 & 0.7 \\ $w$ = 0.94 & 0.9 & 0.26 \\ $w$ = 0.98 & 0.16 & 0.02
\end{tabular}
\end{center}
\end{table}

We have examined the deformations of the black holes for $\alpha > 0$. For simplicity, we have considered
the case of a mass-potential. As pointed out above there exist two black holes corresponding to a choice
of $r_H$ and $w$. We distinguish these solution as belonging to  the $Ex$-branch (connected to extremal black holes) 
and $BS$-branch (connected to boson star).
It turns out that the deformation of a couple of such solutions for $\alpha >0$ leads to two branches of solutions which
terminate into a single solution at some maximal value $\alpha_{max}(w, r_H)$.  
On Fig. \ref{gb_w0995}, we show the dependence of the mass $M$
and of the charge $Q$ on $\alpha$ for $w=0.995$ and for two values of $r_H$ (for instance $r_H = 0.025$ and $r_H = 0.1$).
We have constructed the solutions for several values of $w$ and obtaining the same pattern. Values of $\alpha_{max}$
corresponding to a few choices of $w$ and $r_H$ are given in Table 1. 
For a fixed $w$, the value $\alpha_{max}(w,r_H)$ decreases monotonically with increasing $r_H$. We noticed a strong
decrease with $\alpha$ of the temperature  of the solution corresponding to the $BS$-branch while increasing $\alpha$.

\section{Conclusions}
Very  recently there was a strong interest for astrophysical objects presenting scalar hairs,
see for example \cite{Grandclement:2014msa},
\cite{Meliani:2015zta}, \cite{Vincent:2015xta}, \cite{Troitsky:2015mda}.
This type of research motivate the study of the  possible hair structure occurring around compact gravitating objects.
In many cases, bosons star and/or hairy black holes are continuously connected to vacuum solutions, i.e. they 
loose their hair in some limit of the parameter space. In this respect, the hairy black holes constructed in
\cite{Brihaye:2014nba} are different since their scalar cloud cannot be dissolved by taking
 a special limit; in other words they are disjoined
from the well know Myers-Perry vacuum solutions. This property is, likely, due to a peculiar
non-linear effect in the equations. The question of the persistence of this feature in more general models then
raises naturally.

In this paper, we have extended the results of \cite{Brihaye:2014nba} in two directions:
\begin{itemize}
\item[(i)] We included a self-interaction for the scalar field by mean of a potential possessing a non trivial vacuum manifold.
 We guess that all intermediated potentials between $V_6$ and $V_2$ would lead to similar properties.
\item[(ii)] Extending the gravity sector
by a Gauss-Bonnet term which is believed to encode corrections to Einstein gravity
coming from string theory.
\end{itemize}

The results is that the clouds of the solutions keep their distinguished property in both cases. 
As new features let us
stress that the self-interaction allows for spinning black hole with (in principle) arbitrarily small-but non vanishing-
angular velocity on the horizon, say $w$. 
Only when this parameter become sufficiently small  do the  boson stars constitute classically stable configurations.

In the case of EGB gravity, we have shown that, for all values of $w,r_H$ allowing
for hairy black holes, there exist a couple of corresponding solutions. These two solutions exist up to
a maximal value of the Gauss-Bonnet coupling constant, say $\alpha_{max}$, and converge to a single solution in the limit
$\alpha \to \alpha_{max}$.
\\
{\bf Note added}: While finishing this paper we were aware of \cite{kkmr}
where hairy black holes in four dimensional Einstein-Gauss-Bonnet-dilaton gravity
are constructed.


\end{document}